# Design and Fabrication of a Differential MOEMS Accelerometer Based on Fabry–Pérot micro-cavities

Mojtaba Rahimi, Mohammad Malekmohammad*, Majid Taghavi, Mohammad Noori and Gholam Mohammad Parsanasab

*Abstract*—In this paper, a differential micro-opto-electro-mechanical system (MOEMS) accelerometer based on the Fabry-Pérot (FP) micro-cavities is presented. The optical system of the device consists of two FP cavities and the mechanical system is composed of a proof mass that is suspended by four springs. The applied acceleration tends to move the PM from its resting position. This mechanical displacement can be measured by the FP interferometer formed between the proof mass cross-section and the optical fiber end-face. The proposed sensor is fabricated on a silicon-on-insulator wafer using the bulk micromachining method. The results of the sensor characterization show that the accelerometer has a linear response in the range of ±1g. Also, the optical sensitivity and resolution of the sensor in the static characterization are 6.52 nm/g and 153 µg. The sensor sensitivity in the power measurement is 49.6 mV/g and its resonant frequency is at 1372 Hz. Using the differential measurement method increases the sensitivity of the accelerometer. Based on the experimental data, the optical sensitivity in static mode is two times as high as that of a similar MOEMS accelerometer with one FP cavity.

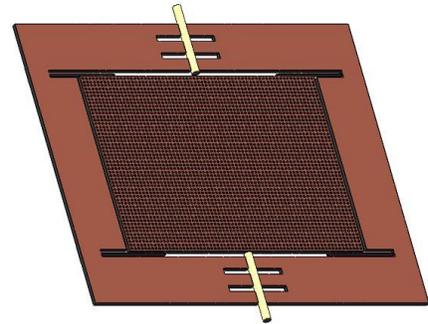

## I. Introduction

MICRO Electro Mechanical Systems (MEMS) accelerometers play an essential role in different applications such as vibration monitoring, inertial navigation, and consumer electronics due to their high sensitivity, small size, lightweight, and easy integration [1]–[5]. MEMS accelerometers can be categorized based on their detection methods are mainly capacitive, piezoelectric, and optical methods [6]–[11]. Among different types of MEMS accelerometers, optical MEMS (MOEMS) accelerometers have intrinsic immunity to electromagnetic interference (EMI) and can be utilized under strong EMI conditions [12]. In the MOEMS accelerometer, applied acceleration moves the proof mass (PM) from its resting position. The mechanical displacement of the PM is measured by monitoring the changes in the output light characteristics such as intensity, frequency, and phase modulations [12]–[15]. The interferometric approach is one of the typical detection methods based on frequency modulation that can provide high sensitivity and wide bandwidth [16]–[18]. Among interferometry-based sensors, Fabry–Pérot (FP)-based MOEMS accelerometers have a high sensitivity to cavity length, a very low cross-axis sensitivity, a simple and integrated configuration and an easy fabrication process [19]–[22]. In FP-based accelerometers, the applied acceleration changes the FP cavity length. The cavity length change leads to an optical spectrum shift that can be used to measure the amplitude and direction of the applied acceleration. Many attempts have been carried out to improve the performance of these accelerometers in terms of sensitivity measurement range, and frequency [22]–[26].

Sensitivity and measurement range are two principal characteristics of an accelerometer. These two characteristics are interdependent in FP-based MOEMS accelerometers due to the repetitive pattern of the FP optical spectrum. The maximum achievable sensitivity of the FP-MOEMS accelerometers is limited in a specific measurement range. Hence, a wide measurement range causes low sensitivity and vice versa In this paper, an FP-MOEMS accelerometer based on a differential measurement method is proposed that can overcome this limitation. This accelerometer has two FP cavities that can double the sensitivity of the sensor in comparison with those of the FP-MOEMS accelerometer with one FP cavity.

The organization of this paper is as follows. In section 2, the working principle of the device is discussed. In addition, the

Mojtaba Rahimi and Mohammad malekmohammad are with the Department of Physics, University of Isfahan, 81746-73441, Isfahan, Iran (m.rahimi@sci.ui.ac.ir, corresponding author: m.malekmohammad@sci.ui.ac.ir).

Majid Taghavi, Mohammad Nori, and Golam Mohammad Parsanasab are with the Integrated Photonics Laboratory, Faculty of Electrical Engineering, Shahid Beheshti University, Tehran, Iran (m_taghavi@sbu.ac.ir, mohammadnooritg@gmail.com, gm_parsanasab@sbu.ac.ir).

design and simulation of the optical, mechanical, and electrical parts of the sensor are described. In section 3, the fabrication process as well as the results of the sensor characterization in both open-loop and closed-loop modes are presented. Finally, the conclusions are given in section 4.

## II. Design and Simulation

### A. Principle of Operation

The working principle of the MOEMS accelerometer is based on the optical measurement of the mechanical displacement caused by acceleration. Fig. 1 shows the schematic of the proposed MOEMS accelerometer. Two FP cavities are the optical sensing systems of the device. The reflective surfaces of each FP cavities are the optical fiber end-face and the PM sidewall. The mechanical structure consists of a suspended PM that can move freely in the sensing axis of the device (Y-axis), and its movements in the two other perpendicular axes are limited. The applied acceleration moves the PM in the opposite direction. The displacement of the PM changes the length of the FP cavities. This change leads to a shift in the optical spectrum of the cavities. Increasing the length of the FP cavity moves the optical spectrum toward a longer wavelength (red shift) and vice versa (blue shift). Therefore, measuring the optical spectrum shift can determine the amplitude and direction of the applied acceleration.

The optical spectrum shift can be measured using two different approaches: spectral monitoring and intensity monitoring. In the spectral method, optical spectrum shift is measured using a spectrum analyzer. In the intensity monitoring method, the intensity variation of a specific wavelength is measured using a photodiode. Due to the limited sampling rate of the optical spectrum analyzers, the first method cannot provide a high dynamic response. Therefore, the intensity monitoring method is more practical in most applications.

Due to the repetitive pattern of the FP optical spectrum, the maximum detectable spectrum shift is limited to the free spectral range (FSR). FSR is the spectral range between two consecutive maximum or minimum of the FP optical spectrum. The optical sensitivity of the MOEMS accelerometer is defined as the ratio of optical spectrum shifts to the applied acceleration ($\Delta\lambda/\Delta a$). For a specific measurement range ($\Delta a$), the maximum achievable sensitivity of the sensor is limited by the FSR of the FP spectrum. To overcome this limitation and achieve higher sensitivity, differential measurement method based on two FP cavities is proposed here. As shown in Fig. 1, the FP cavities (hereafter referred to as "$C_1$" and "$C_2$") are located on the opposite side of the PM. When an acceleration in the +Y direction is applied to the structure, the PM tends to move in the opposite direction. Thus, this acceleration decreases the length of $C_1$, while increasing the length of $C_2$ by the same value and vice versa. Consequently, the measured spectral shift ($\Delta\lambda = \Delta\lambda_1 - \Delta\lambda_2$) doubles the sensitivity as compared to the single FP cavity.

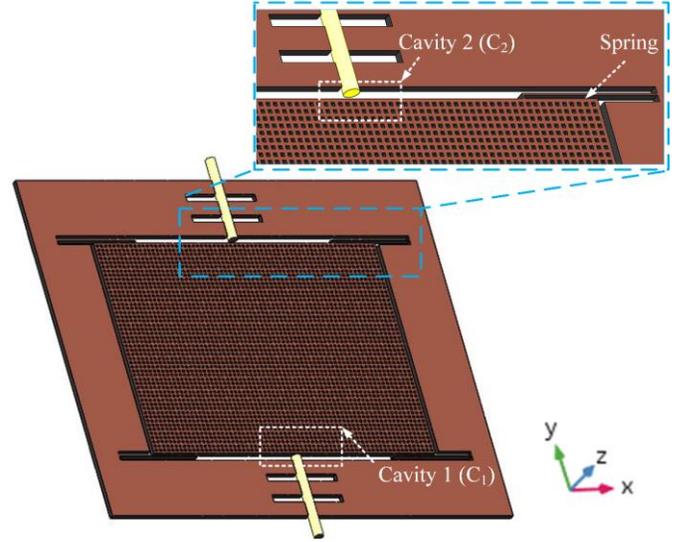

Fig. 1. Mechanical structure of the designed MOEMS accelerometer.

### B. Optical Design

The optical sensing system of the MOEMS accelerometer consists of two FP cavities located on the opposite sides of the PM (C1 and C2). Each FP cavity is formed between the cross-section of the PM and the end of a cleaved single-mode optical fiber. The FP cavities are located along the sensing axis (Y-axis) and at the center of the PM length. The PM displacement changes the length of the FP cavities. When the PM displacement is the +Y direction the length of C2 is decreasing, while the length of C1 is increasing by the same value and vice versa. Hence, two cavities have an equal optical spectrum shift while in the opposite direction.

The finite-difference time-domain (FDTD) method is used to simulate the light propagation in the FP cavities. The characteristics of the FP cavities and parameters used in the optical simulation are mentioned in Table 1. The output spectrum of the FP cavity is shown in Fig. 2-a and the optical sensitivity is 3.27 nm/g as shown in Fig. 2-b. As mentioned in the previous section, intensity monitoring is more practical to measure the optical spectrum shift caused by the applied acceleration. In this method, the intensity variation of the selected wavelength must have linear variation in response to the applied acceleration. The selected wavelength is λ=1550 nm, a commonly used wavelength in optical sensing applications. To determine the linear response range of this wavelength, the intensity has been plotted versus cavity length in Fig. 2-c. The linear response range is highlighted in this figure. It can be seen that the maximum cavity length variation corresponding to the linear response range of the sensor is 260 nm. Since the desired measurement range of the sensor is ±1g, the corresponding mechanical displacement of the PM should be ±130 nm.

Table 1. The parameters used in the optical simulation.

| Parameter | Description | Value |
|---|---|---|
| λ | Source wavelength | 1.5-1.6 µm |
| $D_0$ | Cavity length | 45 µm |
| a | Optical fiber core diameter | 8.2 µm |
| $n_{co}$ | Optical fiber core refractive index | 1.446 |
| $n_{cl}$ | Refractive index of the optical fiber cladding | 1.441 |
| n | Refractive index of the cavity medium | 1 |

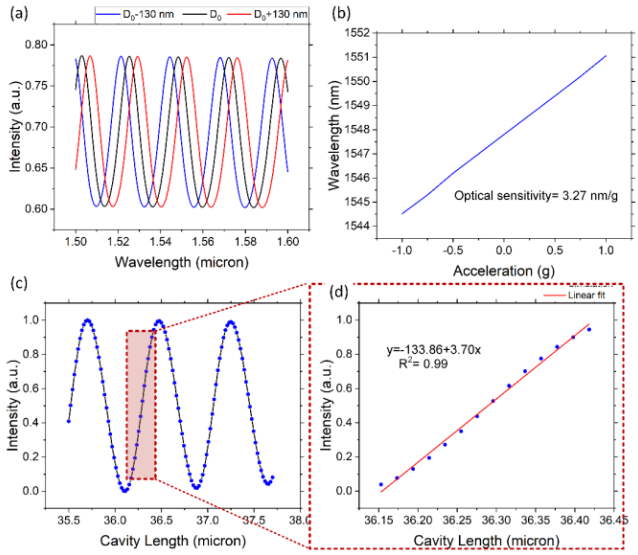

Fig. 2. Intensity variation versus cavity length for the wavelength of 1550 nm and the linear response range of the sensor.

## C. Mechanical Design

The mechanical structure of the device consists of a PM which is suspended by four L-shaped springs. The PM can move freely along the sensing axis, and its movement is restricted to other orthogonal axes to minimize the cross-axis sensitivity. Therefore, the mechanical structure can be considered as a one-dimensional mass-spring system, and its static response is described by the following equation [27].

$$\ddot{y} + \gamma \dot{y} + \omega_0^2 y = -a_y, \qquad \omega_0 = \sqrt{k_y/m} \qquad (1)$$

In this equation, $a_y$ is the applied acceleration, $y$ is the displacement from the rest position, $k_y$ is the effective spring constant of the structure along the sensing axis, $m$ is the mass of the PM, $\gamma$ is the damping coefficient, and $\omega_0$ is the natural resonance frequency. The mechanical displacement of the PM must be matched with the calculated values in the optical design section. To validate this condition for the designed structure, FEM simulation is done using COMSOL Multiphysics. The parameter used in the simulation and the results are summarized in

| Parameters | Description | Value |
|---|---|---|
| L×W | Dimensions of the PM | 4020×3020 μm² |
| l×w | Dimensions of the etch holes | 30×30 μm² |
| Ws | Width of the springs | 15 μm |
| Lx | Length of the springs along the X axis | 1110 μm |
| Ly | Length of the springs along the Y axis | 45 μm |
| t | Thickness | 75 μm |
| ρ | Density of Silicon | 2329 kg/m³ |
| E | Young's modulus | 169 GPa |
| υ | Poisson's ratio | 0.28 |

and Table 3, respectively. The mechanical displacement versus the applied acceleration is shown in Fig. 3. It can be seen that the mechanical sensitivity of the designed structure is 130 nm/g which meets the optical design requirements.

Table 2. The parameters used in the mechanical simulation.

| Parameters | Description | Value |
|---|---|---|
| L×W | Dimensions of the PM | 4020×3020 μm² |
| l×w | Dimensions of the etch holes | 30×30 μm² |
| Ws | Width of the springs | 15 μm |
| Lx | Length of the springs along the X axis | 1110 μm |
| Ly | Length of the springs along the Y axis | 45 μm |
| t | Thickness | 75 μm |
| ρ | Density of Silicon | 2329 kg/m³ |
| E | Young's modulus | 169 GPa |
| υ | Poisson's ratio | 0.28 |

To determine the frequency response of the structure, the four first resonance modes are investigated using the FEM simulation. As shown in Fig. 4, the first resonant mode determining the sensor working bandwidth is at 1368.8 Hz and oscillates completely along the sensing axis (Y-axis). The second resonant mode oscillates along the perpendicular axis (Z-axis), while the third and fourth modes are rotational modes and have the frequency of 8445.1 and 9180.9 Hz, respectively. The frequencies of these modes are well above the working bandwidth of the sensor, so they do not affect the performance of the device.

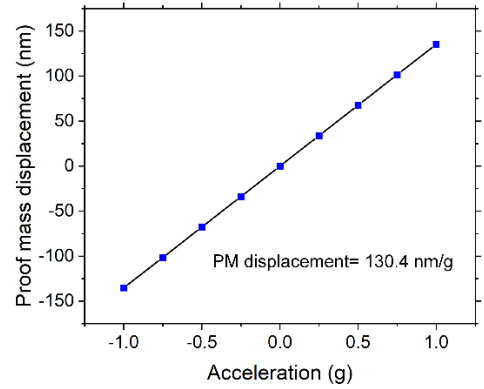

Fig. 3. The PM displacement versus the applied acceleration.

Table 3. The results of the mechanical simulation.

| Symbol | Characteristic | Simulation result |
|---|---|---|
| m | Mass of the PM | 1.71 mg |
| Ky | Spring constant along the Y axis | 128.51 N/m |
| Kx | Spring constant along the X axis | 5.24×10⁶ N/m |
| Kz | Spring constant along the Z axis | 2.48×10⁵ N/m |
| Sy | Mechanical sensitivity along the Y axis | 130.4 nm/g |
| Sx | Mechanical sensitivity along the X axis | 3.20×10⁻³ nm/g |
| Sz | Mechanical sensitivity along the Z axis | 6.76×10⁻² nm/g |

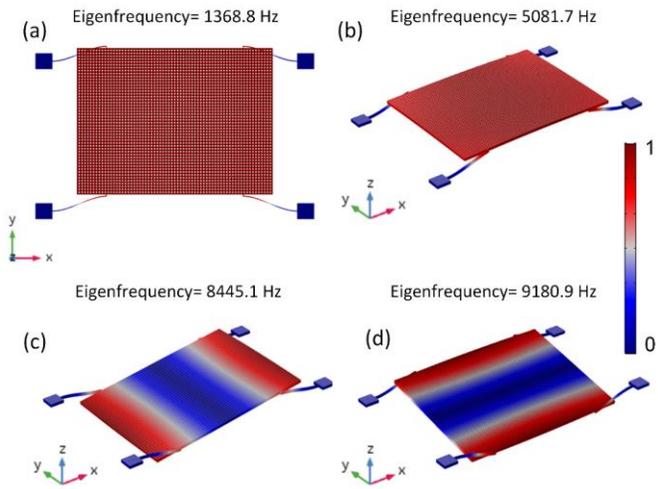

Fig. 4. The four first resonance mode of the structure.

## III. FABRICATION AND CHARACTERIZATION

The sensor is fabricated on a silicon-on-insulator (SOI) wafer by bulk micromachining method. The fabrication process flow is shown in Fig. 5-a. At the first step, the SOI wafer with a 75-micron device layer, a 4-micron oxide layer, and a 470-micron handle layer is cleaned using the RCA procedure. Then, a layer of photoresist with a thickness of 1 micron is spin-coated on the surface of the device layer. The photoresist is patterned using contact mask photolithography. After developing, the masked substrate is ready for the deep reactive ion etching (DRIE) process. In the DRIE process, the structure of the device is formed on the device layer. After removing the photoresist, the oxide layer is removed using HF vapor. The grid structure of etching holes in the PM ensures the accessibility of the HF vapor to the oxide layer in this step. Afterward, the suspended parts of the structure are released. Finally, a gold layer is coated on the sidewalls of the PM to increase the reflectivity of the FP surfaces. Also, this layer prevents unwanted reflections from the inside of the PM.

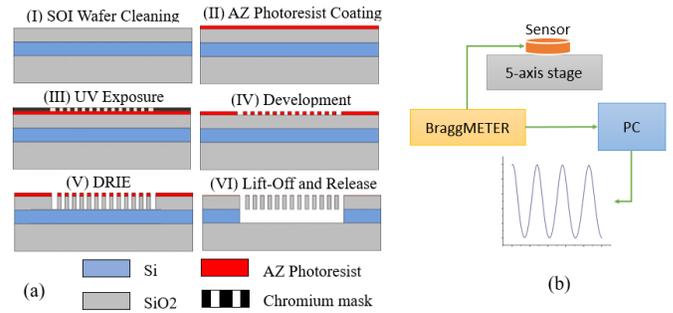

Fig. 5. a) The fabrication process including SOI wafer cleaning, photoresist coating, exposing the photoresist to UV light, photoresist development, DRIE process, and lifting-off the oxide layer; b) the optical fiber assembly setup.

After the fabrication process, the optical fibers are assembled using high-precision positioning stages and a BraggMETER (FS22 Industrial BraggMETER). The optical setup used in fiber assembly is shown in the Fig. 5-b. As previously mentioned, each FP cavity is formed between the PM sidewall and the end of a single-mode optical fiber. The optical fiber is inserted into micromachined grooves using a 5-axis stage. The other end of the optical fiber is connected to BraggMETER which send the light with a wavelength of 1500- 1600 nm to the FP cavity. In the optical design section, the FP cavity length was selected as 45 microns. This length is equivalent to the formation of about 4 peaks in BraggMETER wavelength range. Therefore, by monitoring the optical spectrum, the length of the FP cavity is adjusted using the high-precision stage. Finally, the optical fiber is fixed by a UV cure adhesive. This process is repeated for another FP cavity. The SEM image of the fabricated device is shown in Fig. 6.

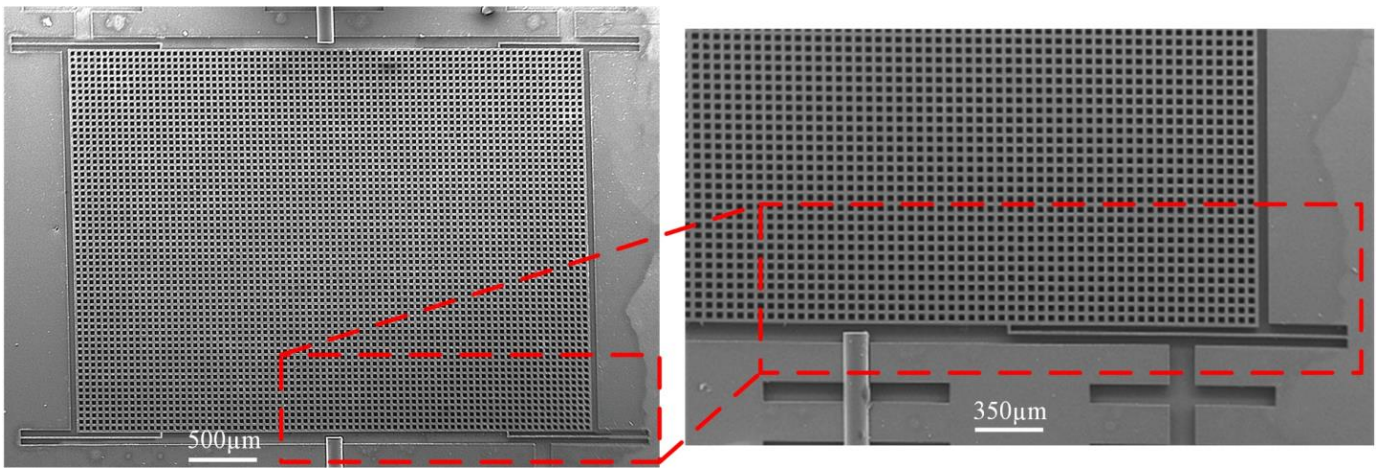

Fig. 6. SEM image of the fabricated device.

## IV. EXPERIMENTS AND RESULTS

### A. Static characterization

To determine the sensitivity of the sensor, the fabricated MOEMS accelerometer is placed on top of a turning table. The schematic of the experimental setup is shown in Fig. 7. The sensor is placed on top of a turning table. By rotting the turning table with a stepper motor, a static acceleration is applied to the sensor. The amplitude of the applied acceleration is a function of θ as shown in Fig. 7 (a=gsinθ). Therefore, changing θ from -90° to +90°, can provide acceleration in the range of -1g to +1g. Light in the wavelength ranging from 1500 nm to 1600 nm is sent to the FP cavities through two outputs of BraggMETER. The reflected interference spectrum is measured by BraggMETER and monitored on the PC. Also, A commercial MEMS accelerometer (ADXL 202C) is used as a reference accelerometer.

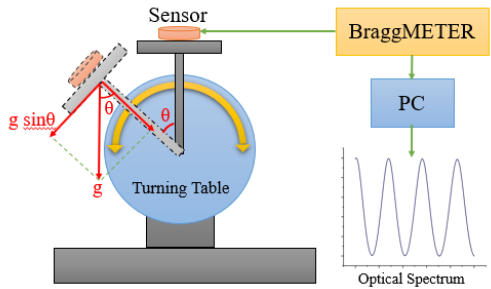

Fig. 7. Schematics diagram of experimental setup for static characterization.

The optical spectrum shift of $C_1$ and $C_2$ in response to the static acceleration is shown in Fig. 8-a and Fig. 8-c, respectively. While the optical spectrum of $C_1$ moves toward a longer wavelength (red shift), the optical spectrum of $C_2$ moves to a shorter wavelength (blue shift). The spectral shift of $C_1$ and $C_2$ versus applied acceleration is shown in Fig. 8-b and Fig. 8-d, respectively. The values and the error-bars in the graphs are obtained by five times repetition of the experiments. It is noticeable that both FP cavities have a linear response in the ±1g measurement range, and their optical sensitivity is 3.26 nm/g. This value is in a good agreement with the simulation results. The optical spectrum shift of $C_1$ and $C_2$ have the same absolute value but different signs. Hence, the total optical spectrum shift can be considered as $\Delta\lambda = \Delta\lambda_1 - \Delta\lambda_2$, and the optical sensitivity of the sensor is 6.52 nm/g. This optical sensitivity is two times as high as that of the MOEMS accelerometer with one FP cavity in our previous work [28]. Also, the sensor resolution is defined as the ratio of BraggMETER precision to the optical sensitivity of the sensor. The BraggMETER used in the experiments has a precision of 1 picometer, so the sensor resolution is 153 µg based on the following equation.

$$resolution = \frac{BraggMETER\ resolution}{Sensitivity} \quad (2)$$

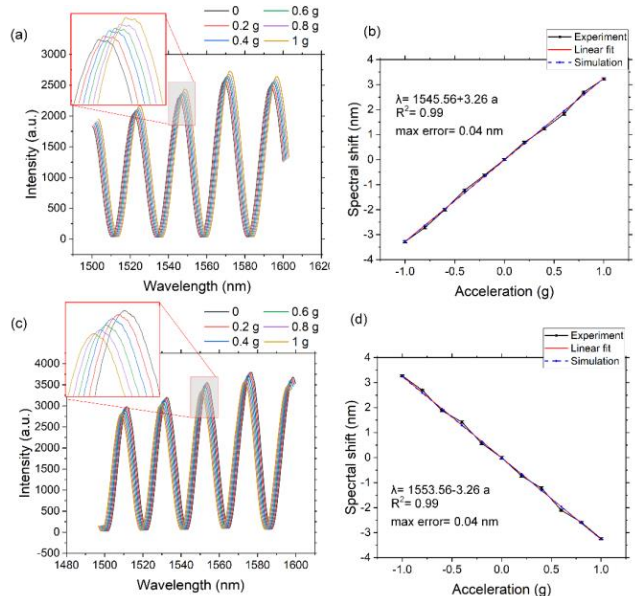

Fig. 8. a) $C_1$ optical spectrum shift due to the applied acceleration, b) The spectral behavior of $C_1$ versus the applied acceleration, c) the optical spectrum shift of $C_2$, d) The spectral behavior of $C_2$ versus the applied acceleration.

## B. Dynamic characterization

To investigate the dynamic response of the sensor, a shaker that can apply horizontally sinusoidal acceleration is used instead of the turning table in the previous test. The experiment setup is shown in Fig. 8, schematically. Light from a broad band laser (S5FC1550S-A2-SM Benchtop SLD Source) passes through an isolator and a coupler and enters the tunable optical filters (Agiltron-FOTF-029121112). The reflected light from each cavity is sent to photodiodes (PDA10CS-EC Thorlabs) through couplers. The optical filters are used to set the laser wavelength between two successive maximum and minimum of the optical spectrum. In this experiment, the intensity monitoring is used instead of spectral monitoring. So, a specific wavelength should be selected for the cavity input light. To achieve the full linear response range (the highlighted region in Fig. 2-c), the central wavelength between two consecutive peak and deep of the optical spectrum should be selected. This central wavelength is dependent on the FP cavity length. Because there is a small difference between the length of the cavities, they have different central wavelengths. These wavelengths are selected by monitoring the cavities' spectrum in BraggMETER. These wavelengths are set to 1547.2 nm and 1551.9 nm for C1 and C2, respectively.

For the dynamic characterization of the sensor, the shaker applies the sinusoidal acceleration with the amplitude of 0.25g to 1g and the frequency of 150 Hz. As shown in the Fig. 10-a and Fig. 10-c, the output voltage of the photodiodes follows the applied acceleration and oscillates with the same frequency. The output voltage of the photodiode versus the applied acceleration for C1 and C2 are shown in Fig. 10-b and Fig. 10-d, respectively. The sensitivity of each cavity is 24.8 mV/g, and the sensor has a linear response in the measurement range. As discussed in the previous section, using the differential measurement method, the total sensitivity of the sensor is 49.6 mV/g which is twice higher than that of a same MOEMS accelerometer with one FP cavity. Also, the response of the sensor and the reference accelerometer are compared in Fig. 11-a when a 1g acceleration with a frequency of 150 Hz is applied.

The frequency response of the sensor is investigated by sweeping the frequency of the shaker in the range of 500 Hz to 1700 Hz. The response of the sensor is shown in Fig. 11-b. The resonance frequency of the sensor is at 1372 Hz which is slightly different from the obtained value in the simulation. This difference can be attributed to the slight variation of dimensions in the fabrication process.

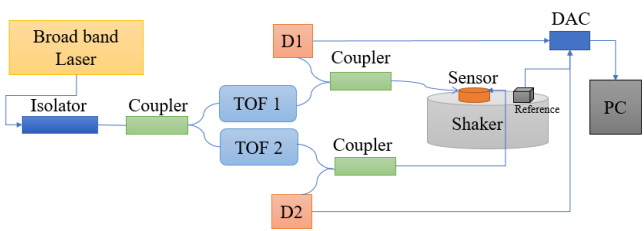

Fig. 9. The schematic of the experimental setup for the dynamic characterization of the sensor. $D_1$ and $D_2$ are photodiodes and TOFs are tunable optical filters.

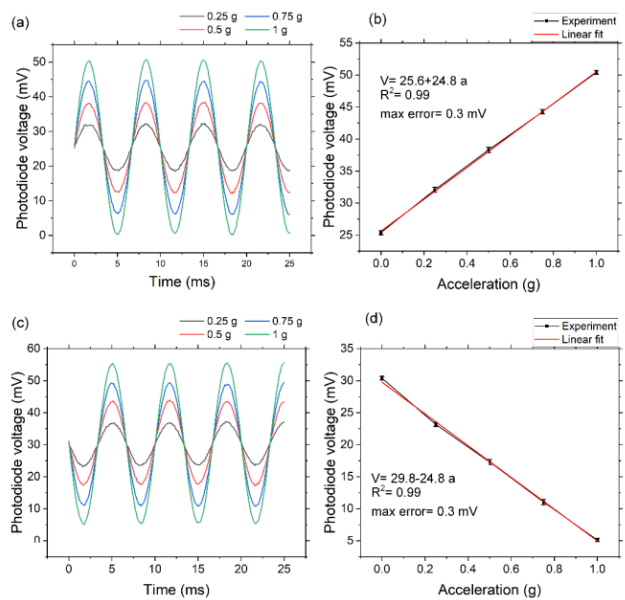

Fig. 10. a) The dynamic response of $C_1$ to four different accelerations with the same frequency of 100 Hz, b) linearity of the respons of $C_1$, c) the response of $C_2$ to the dynamic accelerations, d) linearity of the respons of $C_2$.

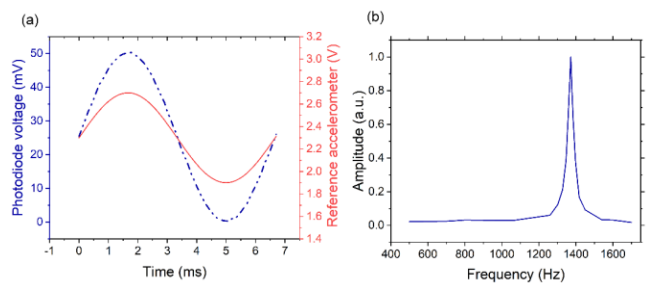

Fig. 11. a) The responses of the sensor and reference accelerometers at 150 Hz, b) the frequency response of the sensor.

## C. Analysis and comparison

To compare the performance of the presented MOEMS accelerometer with those of other sensors in this field, the main characteristics of the sensor and three recent accelerometers are summarized in Table 4. Considering the sensitivity and the measurement range, the sensitivity of the fabricated accelerometer is much higher than those reported in previous works. This is a significant characteristic of our sensor. Furthermore, the dimensions of the fabricated MOEMS accelerometer are in the millimeter scale and its structure is completely integrated, whereas the accelerometers in [22],[29] are bulky and have higher volumes and masses. Moreover, the fabrication process of the sensor is based on a straightforward bulk micromachining method which facilitates the mass production of the device.

Table 4. The comparison of the main characteristics of the fabricated sensor with those of the accelerometers in three recent works.

| Characteristics | This work | 2021 [30] | 2020 [22] | 2019 [29] |
|---|---|---|---|---|
| Measurement range | ±1 g | ±1 g | ±2.5 mg | ±5.2 g |
| Optical sensitivity (Static Mode) | 6.52 nm/g | 3.22 nm/g | - | 1.63 nm/g |
| Resolution | 153 µg | 313 µg | 0.3 µg | - |
| Resonance frequency | 1372 Hz | 1382 Hz | 89 Hz | 463 Hz |
| Footprint of the PM | 4 × 3 mm² | 2× 2 mm² | 5× 5 mm² | 17× 10 mm² |
| Sensing structure | Fabry-Pérot cavity (Integrated) | Fabry-Pérot cavity (Integrated) | Fabry-Pérot cavity (Bulk) | Blazed Grating (Bulk) |

## V. CONCLUSION

In this paper, the design and fabrication of a MOEMS accelerometer based on FP micro-cavities was reported. The sensor consisted of a PM suspended by four springs. The displacement of the PM was measured by two FP micro-cavities. The read-out system of the MOEMS accelerometer was based on differential measuring the output of the two optical signals. This approach could overcome the limitation to achieving high sensitivity in the FP-based MOEMS accelerometers and provide the optical sensitivity two times as high as the same MOEMS accelerometer with one FP cavity. The sensor was fabricated by a straightforward bulk micromachining method. The dimensions of the fabricated MOEMS accelerometer were in the millimeter scale and its structure was completely integrated. The fabricated device is characterized in the range of ±1 g. According to experimental data, the sensor provided the optical sensitivity of 6.52 nm/g in the static mode and the sensitivity of 49.6 mV/g in the dynamic characterization. The resonant frequency of the sensor was 1372 Hz in the range of ±1g.